\documentclass[twocolumn, linenumber]{aastex63}

\newcommand{\msun}{\ensuremath{{\rm M}_\odot}}

\graphicspath{{./}{figures/}}

\usepackage[perpage]{footmisc}
\usepackage{colortbl}
\usepackage{lineno}
\usepackage{appendix}
\definecolor{Gray}{gray}{0.9}

\begin{document}

\title{Ain't No Mountain High Enough: Semi-Parametric Modeling of LIGO-Virgos Binary Black Hole Mass Distribution}

\author{Bruce Edelman}
\email{bedelman@uoregon.edu}
\affiliation{Institute  for  Fundamental  Science, Department of Physics, University of Oregon, Eugene, OR 97403, USA}

\author{Zoheyr Doctor}
\affiliation{Institute  for  Fundamental  Science, Department of Physics, University of Oregon, Eugene, OR 97403, USA}
\affiliation{Center for Interdisciplinary Exploration and Research in Astrophysics (CIERA), Department of Physics and Astronomy, Northwestern
University, Evanston, IL 60201, USA}

\author{Jaxen Godfrey}
\affiliation{Institute  for  Fundamental  Science, Department of Physics, University of Oregon, Eugene, OR 97403, USA}

\author{Ben Farr}
\affiliation{Institute  for  Fundamental  Science, Department of Physics, University of Oregon, Eugene, OR 97403, USA}

\begin{abstract}
We introduce a semi-parametric model for the primary mass distribution of binary black holes (BBHs) observed with gravitational waves (GWs) that applies a cubic-spline perturbation to a power law. We apply this model to the 46 BBHs included in the second gravitational wave transient catalog (GWTC-2). The spline perturbation model recovers a consistent primary mass distribution with previous results, corroborating the existence of a peak at 35\msun ($>97\%$ credibility) found with the \textsc{Powerlaw+Peak} model. The peak could be the result pulsational pair-instability supernovae (PPISNe). The spline perturbation model finds potential signs of additional features in the primary mass distribution at lower masses similar to those previously reported by \citet{Tiwari_2021_b}. However, with fluctuations due to small number statistics, the simpler \textsc{Powerlaw+Peak} and \textsc{BrokenPowerlaw} models are both still perfectly consistent with observations. Our semi-parametric approach serves as a way to bridge the gap between parametric and non-parametric models to more accurately measure the BBH mass distribution. With larger catalogs we will be able to use this model to resolve possible additional features that could be used to perform cosmological measurements, and will build on our understanding of BBH formation, stellar evolution and nuclear astrophysics.
\end{abstract}

\section{Introduction} \label{sec:intro}
The LIGO-Virgo Collaboration's second catalog of compact object mergers has shown that the universe is teeming with colliding compact objects with a variety of masses and spins \citep{aligo_detectors}. In contrast to the 11 sources reported in the first LIGO-Virgo Collaboration (LVC) catalog \citep[GWTC-1][]{GWTC1}, the second catalog \citep[GWTC-2][]{gwtc2} contains 50 sources, enabling a deeper look into the formation environments of compact object binaries. 
The sources in GWTC-2 include two binary neutron stars (BNSs) \citep{170817disc, 190425disc}, 46 binary black holes (BBHs), and two neutron star black hole (NSBHs) candidates \citep{190814disc}. The 46 confirmed BBHs observed in GWTC-2 include the first clear evidence of an asymmetric mass binary, potentially the least massive black hole known, and the most massive stellar mass black hole to date \citep{190412_disc, 190814disc, 190521_disc, 190521_astro_imp}. With this large catalog of BBH mergers, one can now begin to robustly infer the properties of the astrophysical BBH distribution in addition to each individual event properties \citep{o1o2_pop, o3a_pop}. 

Prior to the release of GWTC-2, the inferred mass distribution for the more massive (primary) components in mergers was thought to be consistent with a declining power law that cuts off at $\sim45\msun$ \citep{o1o2_pop, Fishbach_2017}. When analyzing the BBH primary mass distribution including events in GWTC-2, \citet{o3a_pop} found that a truncated power law is no longer consistent with the additional observations. The primary mass distribution was found to have some feature at $\sim35$--$40\,\msun$, which was best described by either a break to a steeper power law or a with the addition of a peak. The presence of a peak in the primary mass distribution in this mass range is not surprising: it would be expected if we are witnessing effects of pulsational pair-instability supernovae (PPISNe) \citep{Talbot_2018}. Massive stars that are too light to be fully disrupted by a pair-instability supernova (PISN) can shed large amounts of mass in a series of explosive pulsations before collapsing to a black hole \citep{Woosley_2017, Woosley_2019, Farmer_2019}. This process leads to a wide range of initial stellar masses that map onto remnant black holes with masses $30\,\msun \leq m_\mathrm{BH} \leq 45\,\msun$ \citep{Belczynski_2016, Marchant_2019, Stevenson_2019}. GWTC-2 also includes more massive binaries than previously observed, most notably GW190521 \citep{190521_disc, 190521_astro_imp}. Both component black holes of GW190521 have masses that pose a challenge to the theoretical prediction that pair-instability (PI) would forbid isolated stellar evolution from creating remnant black holes with masses from $\sim 50$--$125\,\msun$ \citep{PISN_Woosley, Heger_2002, Heger_2003, Spera_2017}. There is some evidence that GW190521 could be a mass gap straddling binary or the result of other physical processes that get around the conflict with PISN theory \citep{Fishbach_2020, Nitz_2021, estell190521, cruzosorio2021gw190521, Sakstein_2020, gayathri2020gw190521, Farrell_2020, Secunda_2020, Edelman_2021_HOLES}. However, the presence of these high mass component black holes could also point towards there being a contribution to the observed population of BBHs detected by LIGO/Virgo, that formed in a way that avoids PI. These formation possibilities include hierarchical mergers in dense stellar environments, relativistic accretion onto heavy BHs in active galactic nuclei disks, isolated binary evolution of low-metallicity Population II stars, or even the presence of new physics beyond the standard model \citep{Rodriguez_2019, Doctor_2020, Kimball_genealogy, kimball2020evidence, doctor2021black, McKernan_2020, Belczynski_2020, Croon_newphysics, mapelli2020hierarchical}. 

Incorrectly inferring the BBH mass distribution has been shown to significantly bias both estimates of merger rates and the stochastic gravitational wave background amplitude \citep{Talbot_2018}. Additionally the effects of PI can imprint features onto the mass distribution such as a high mass cutoff in the mass distribution (PISN), or a possible a pileup of mergers at masses just below the cutoff (PPISN). Resolving either of these features can provide a mass scale, calibrated across cosmic time, that enables measurements of the redshift-luminosity-distance relation to infer cosmological parameters \citep{Farr_2019HUB}. As catalogs of GWs from BBHs grow in size \citep{prospectso3o4o5}, we will be able to infer the BBH mass distribution with greater fidelity to determine if there is presence of additional structure beyond a power law. Such structure could yield insights about the nature of what environments BBHs form in and how they are connected to the rich fields of stellar evolution and nuclear astrophysics \citep{Zevin_2017, Farmer_2019, Farmer_2020, Ng_2021}.

\begin{figure}
    \centering
    \includegraphics[width=0.475\textwidth]{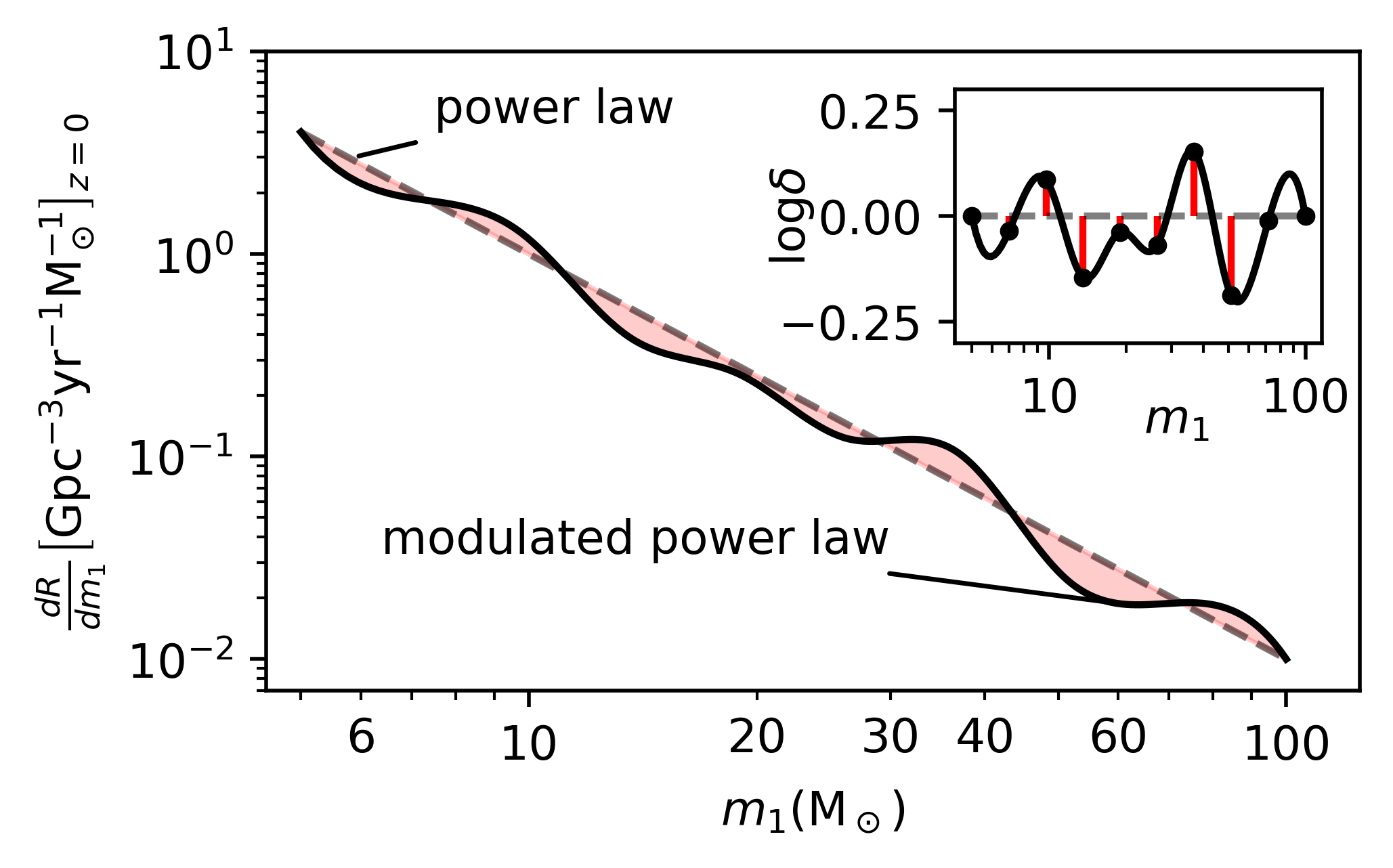}
    \caption{Sketch description of the spline perturbation primary mass model. The inset shows the interpolated cubic spline perturbation function for the plotted modulated power law.}
    \label{fig:model_sketch}
\end{figure}

Bayesian non-parametric models provide a useful data-oriented approach to modeling when one has little information or prior knowledge about the structure of a set of data. These approaches provide little to no constraints on the functional form imposed by the model and instead use very flexible functions that have large prior support for a wide variety of unknown densities. Non-parametric modeling has been widely applied across different areas of GW Astrophysics, including modeling deviations from GR waveform models, modeling the noise power spectrum of detectors, modeling the calibration of the detectors, and creating surrogate models for faster waveform execution \citep{Edelman_2021, Littenberg_2015, Vitale_2021, Doctor_GPR}.

In this work, we approach the mass spectrum from a data-driven perspective, using a semi-parametric method rather than the low-dimensional parametric models used in \citet{o1o2_pop, o3a_pop}. Our semi-parametric method is complementary to both parametric and fully non-parametric approaches \citep{Mandel_2016, Tiwari_2021_a} by incorporating a simple parametric description of the large-scale structure (i.e. a power law) with an additional non-parametrically modeled component on top. This approach can aid in searching for generic deviations to the underlying parametric descriptions that could be the result of astrophysical processes. Since non-parametric approaches make few assumptions on the form that the underlying distributions may take, our model minimizes biases to the structure such deviations could take. We expect a large fraction of stellar mass BHs to form at the end of life of massive stars, which motivates our choice of a power law form of the BBH primary mass distribution following a similar functional form to the stellar initial mass function \citep{Kroupa}. We therefore reconstruct the primary mass distribution with the \textsc{Truncated} power law model \citep{Fishbach_2017, o3a_pop}, in which we modulate with a non-parametric perturbation. This method takes advantage of using a simple parametric form to capture the majority of the structure in the primary mass distribution while the perturbation function can find data-informed deviations from the power law. In Section \ref{sec:spline_model} we describe our semi-parametric perturbation population model, and in Section \ref{sec:results} we present and discuss the inferred properties of the primary mass distribution when analyzing all 46 BBHs in GWTC-2. Finally, we explore possible interpretations of our results and conclude in Section \ref{sec:conclusion}.

\begin{figure*}
    \includegraphics[width=\textwidth]{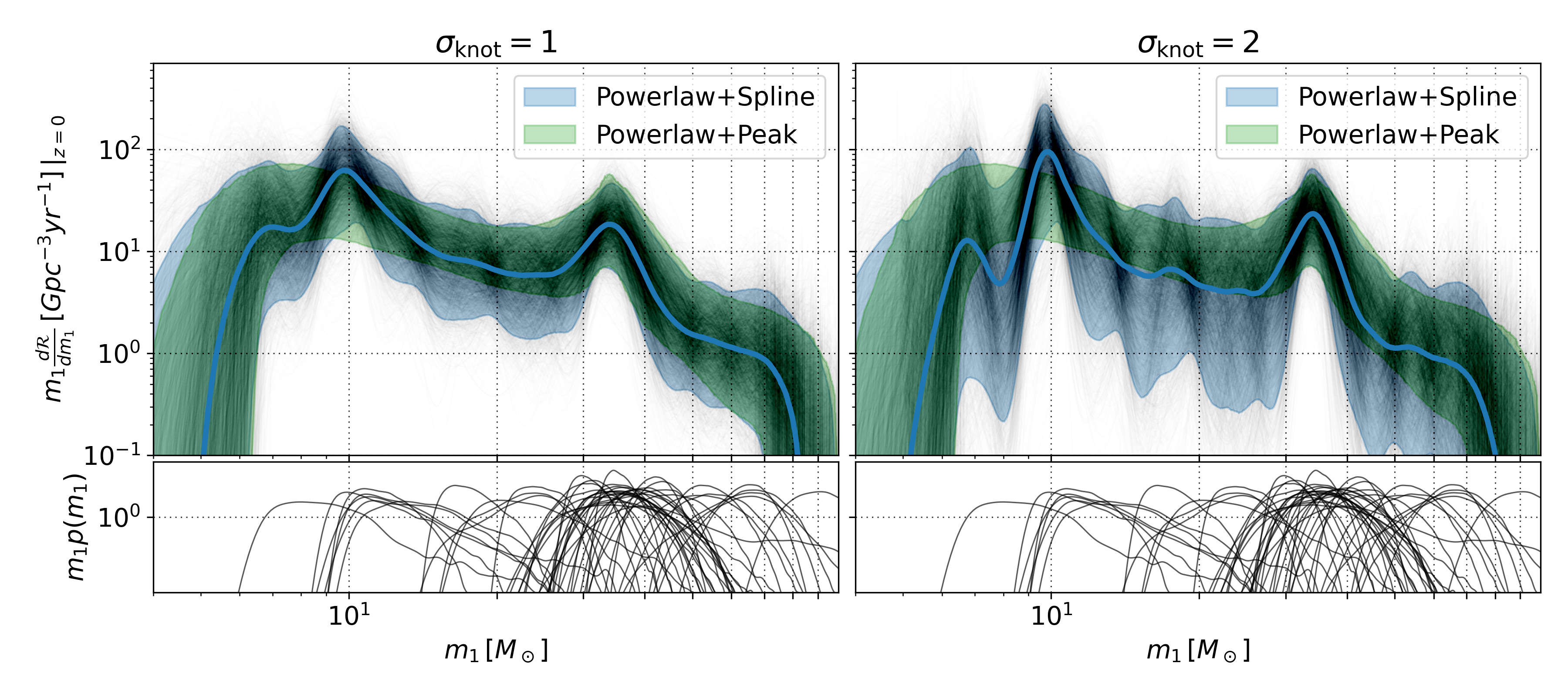}
    \caption{We plot the differential merger rate, $\frac{d\mathcal{R}}{d\log m_1} = m_1\frac{d\mathcal{R}}{m_1}$ as a function of primary mass (top row) for the combined spline model marginalized over the 10, 15, and 20 knot models in blue. The solid line shows the median while the shaded region shows the 90\% credible interval and the 90\% credible interval found from the \textsc{Powerlaw+Peak} model in green. The black traces show 1000 draws from the combined spline model posterior and we plot kernel density estimates (KDEs) of the posterior samples of primary source frame mass for each of the 46 BBHs in GWTC-2. We plot $m_1 p(m_1)$ on a log-scaled y-axis with a Gaussian KDE approximating $p(m_1)$ for each event. These posterior samples are not re-weighted to a population and come directly from the accompanying data release to \citet{gwtc2}.}
    \label{fig:marged_rate}
\end{figure*}

\section{Spline Perturbation Model} \label{sec:spline_model}
We use a hierarchical Bayesian inference framework to infer the properties of the astrophysical distribution of BBHs that incorporates software injections to estimate selection effects \citep{Mandel_2019, Farr_2019, vitale2021inferring}. This procedure is described in detail in Appendix \ref{sec:hierarchical_inference}. In order to capture both the overall trends and any sharper features that may be in the primary mass distribution, we modulate a base parametric hyper-prior on primary mass, $p(m_1|\Lambda)$, by a highly flexible perturbation function -- in this case, a cubic spline. We choose the simplest of previously used parametric models as our underlying mass distribution, $p(m_1 | \Lambda)$, which is described by a power law in both primary mass and mass ratio with a sharp low and high mass cutoff \citep{o1o2_pop, Fishbach_2017, o3a_pop}. This model was referred to as the \textsc{Truncated} model in \citet{o3a_pop}. While the \textsc{Truncated} model alone does not describe GWTC-2 well \citep{o3a_pop}, it captures the majority of the large-scale structure found in the primary mass distribution. For our underlying description, we extend the \textsc{Truncated} model to allow for a tapering of the distribution at low masses following the same form used for the \textsc{Powerlaw+Peak} model described in \citet{Talbot_2018} and \citet{o3a_pop}. Figure \ref{fig:model_sketch} shows an illustration of our spline perturbation model on top of a power law without any mass cutoffs or tapering. We multiplicatively apply perturbations to the underlying distribution as:
\begin{equation}
    p_\mathrm{spline}(m_1 | \Lambda, \{m_i,f_i\}) = k*p(m_1 | \Lambda)\exp(f(m_1;\{m_i,f_i\}))
\end{equation}

\noindent In the above equation, $k$ is a normalization factor found by numerically integrating $p_\mathrm{spline}(m_1 | \Lambda,\{m_i,f_i\})$ over the entire range of primary masses, and $f(m_1;\{m_i,f_i\})$ is the perturbation function modeled as a cubic spline that is interpolated between $n$ knots placed in $m_1$ space. These knots are denoted by their locations in mass space, $\{m_i\}_{i=1}^n$, and their heights at each knot, $\{f_i\}_{i=1}^n$. For readability, we hereafter drop explicit dependence of $f$ on $\{m_i,f_i\}$ unless needed.

\begin{figure*}[ht!]
    \centering
    \includegraphics[width=\textwidth]{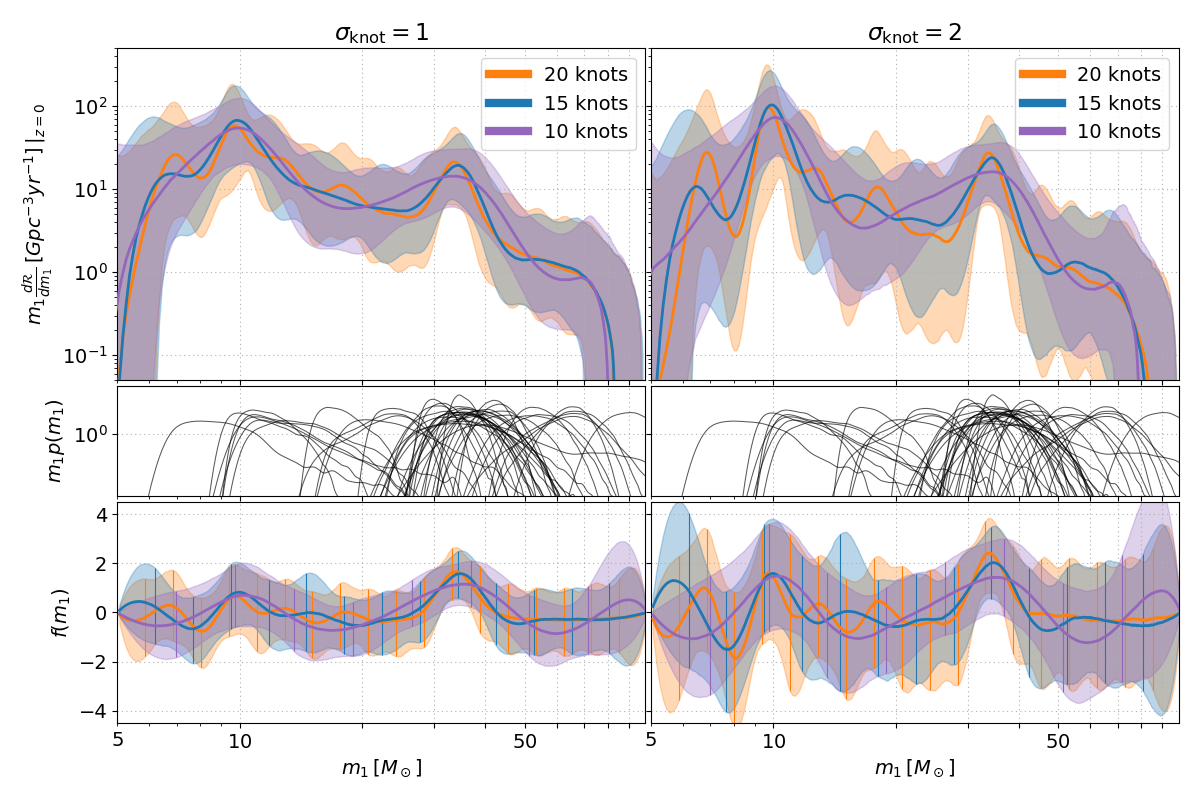}
    \caption{The median (solid) and 90\% credible interval (shaded) of the inferred differential merger rate density as a function of primary mass (top row) with spline models with 10 (purple), 15 (blue), and 20 (orange) knots. We show the conservative prior case ($\sigma_\mathrm{knot}=1$) in the left column and the wide prior case ($\sigma_\mathrm{knot}=2$) on the right. The middle row shows kernel density estimates (KDEs) of the posterior samples of primary source frame mass for each of the 46 BBHs in GWTC-2. We plot $m_1 p(m_1)$ on a log-scaled y-axis with a Gaussian KDE approximating $p(m_1)$ for each event. These posterior samples are not re-weighted to a population and come directly from the accompanying data release to \citet{gwtc2}. The bottom row shows the median (solid) and 90\% credible intervals (shaded) of the inferred perturbation function, $f(m_1)$, for each choice of $n$, with the vertical lines showing the locations of the spline knots.} 
    \label{fig:primary_mass}
\end{figure*}

We fix the locations of each knot to be linear in $\log\,m_1$ space from $5$--$100\,\msun$ and restrict the perturbations to zero at the minimum and maximum knots. With these restrictions our spline model adds $n-2$ extra hyper-parameters to the underlying primary mass model we are perturbing, one for each of the inner knots' heights. We log-space the knots and perturb our underlying model with the multiplicative factor, $\exp(f(m_1))$, to reflect the wide range in orders of magnitude of the underlying power law. We then impose Gaussian priors on the knot heights $\{f_i\}$ centered at 0 and with standard deviations, $\sigma_\mathrm{knot}$. Our model then has two specifications which control the resolution ($n$) and the magnitude ($\sigma_\mathrm{knot}$) of perturbations the model is sensitive to. We discuss the effect of changing these model settings on our prior assumptions and motivate the particular choices we made for this work in Appendix \ref{model-comparison}.

In addition to the primary mass distribution, we simultaneously fit for the mass ratio and redshift distributions, without any spline perturbations applied. We apply a power law distribution for the mass ratio as $p(q | m_1, m_\mathrm{min}, \beta_q) \propto q^{\beta_q} \Theta(qm_1 - m_\mathrm{min}) \Theta(m_1 - qm_1)$, with $\Theta$ denoting the Heaviside step function that ensures $m_2$ is within the range [$m_\mathrm{min}$, $m_1$]. We then fit for the evolution of the merger rate with redshift also with a power law such that $p(z|\lambda) \propto \frac{dV_c}{dz}\frac{1}{1+z}(1+z)^\lambda$, where $dV_c$ is the co-moving volume element \citep{Fishbach_2018redshift, o1o2_pop, o3a_pop, hogg_cosmo}. We do not fit for a population prior on the BBH spins, and assume the spin prior used for individual event parameter estimation in \citet{gwtc2}, which is uniform in component spin magnitudes and isotropic in component spin orientations. We enumerate each of the model's hyper-parameters and corresponding hyper-prior distributions used in this work in Table \ref{tab:model_params}. 

\begin{figure*}
    \centering
    \includegraphics[width=\textwidth]{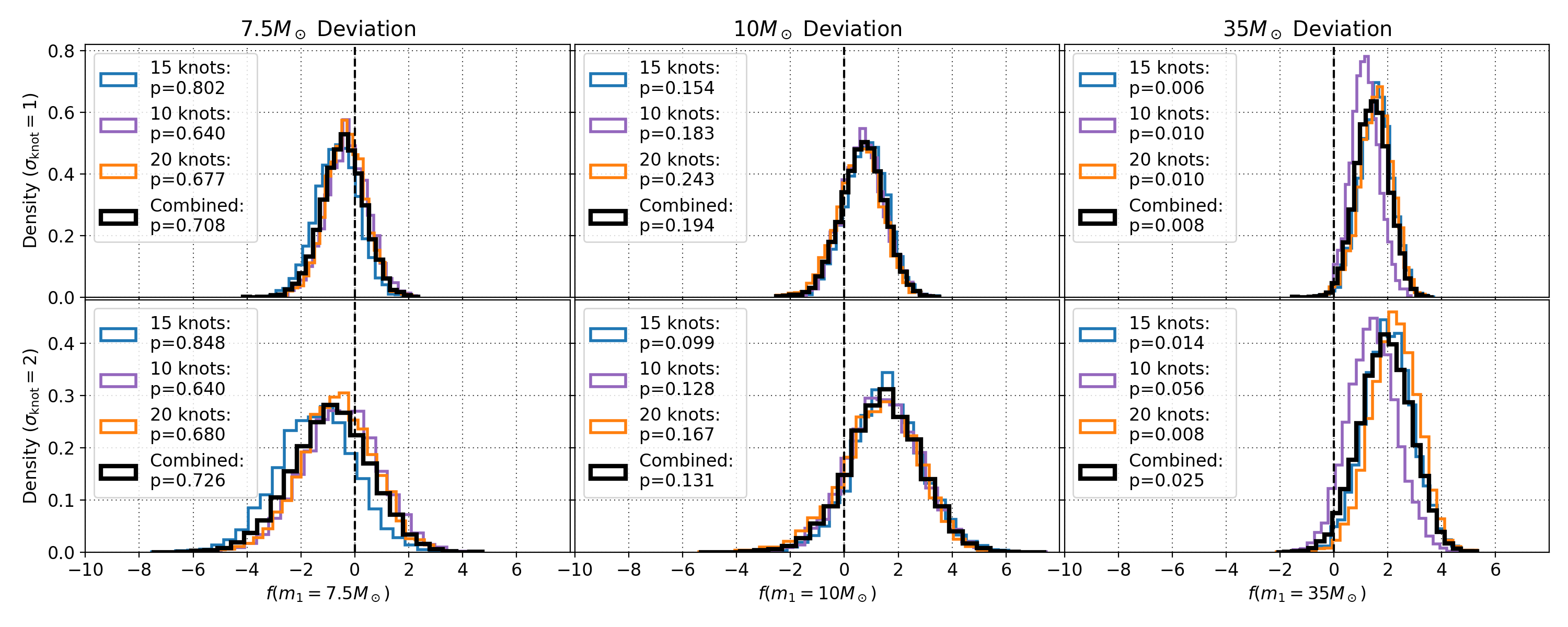}
    \caption{The posterior distribution of $f(m_1)$ at sliced at the three most apparent inferred perturbations in the posterior which roughly lie at $\sim7.5\msun$ (left column), $\sim10\msun$ (middle column), and $\sim35\msun$ (right column). We show the posteriors for 10 (purple), 15 (blue), and 20 (orange) knots and for both cases of prior width: $\sigma_\mathrm{knot} = 1$ (top row) and $\sigma_\mathrm{knot} = 2$ (bottom row). We additionally show the result when combining the models (weighted by their marginal likelihoods) across the three choices number of nodes in black. We report the quantile in which $f=0$ falls for each of the models and perturbation regions knots in each legend.} 
    \label{fig:peak_sigs}
\end{figure*}

\section{Results} \label{sec:results}
\subsection{Astrophysical BBH Primary Mass Distribution}
We use a catalog containing each of the 46 BBH mergers reported in \citet{gwtc2} in which we perform a hierarchical Bayesian analysis to infer the astrophysical mass spectrum and merger rate evolution with redshift, as described in Appendix \ref{sec:hierarchical_inference}. We perform multiple iterations of our semi-parametric model with different numbers of knots and both a ``conservative" ($\sigma_\mathrm{knot}=1$) and ``wide" ($\sigma_\mathrm{knot}=2$) prior width on the knots. For both cases of prior width we additionally do a post hoc ``marginalization'' over the number of knots by combining posterior draws weighted according to the ratios in marginal likelihoods. Explicitly we take $\left\lfloor{N_\mathrm{min}\frac{\mathcal{Z}_n}{\mathcal{Z}_\mathrm{max}}}\right\rfloor$ samples from each inference where $N_\mathrm{min}$ is the minimum number of samples from each posterior, $\mathcal{Z}_n$ the marginal likelihood of inference with $n$ knots, and $\mathcal{Z}_\mathrm{max}$, the maximum marginal likelihood of the combined posteriors. In Figure \ref{fig:marged_rate} we plot the posterior merger rate density as a function of primary mass (top row) for our combined spline model (combining 10, 15 and 20 knot models), compared to the \textsc{Powerlaw+Peak} model. The most prominent feature in the primary mass distribution is the apparent peak at $\sim 35\msun$, similar to the peak found at the same mass by the \textsc{Powerlaw+Peak} model \citep{o3a_pop,Talbot_2018}. 

In addition to the peak at $\sim 35\msun$, there are signs of additional features --- albeit less significant --- at lower masses. We find signs of an inflated rate of mergers with primary masses $\sim 10\msun$ and reduced rate around $\sim 7.5\msun$, when compared with the power law structure. The model is less certain about the low-mass features as there are only a few events with support for $m_1 < 10\,\msun$. The dearth of observed low mass BBHs, combined with their small sensitive volume, significantly inflates our uncertainty at the low end of the mass distribution. In Figure \ref{fig:primary_mass}, we show the results inferred by each of the 10, 15, and 20 knot spline models individually, showing that the spline model consistently finds common features regardless of the number and location of knots. The combined spline model based on the marginal likelihoods is mostly comprised of the 15 knot result because the ratios of marginal likelihoods favor 15 knots over 20 at 2:1 odds and 15 over 10 at 3:1. The spline models are best constrained in the regions of over/under densities discussed above and much less certain (prior dominated) in regions between the features where the parametric component (i.e. power law) can fully explain the trend. The bottom row of Figure \ref{fig:primary_mass} shows the perturbation function, $f(m_1)$, inferred from the different spline models. While there are some differences between knot choices due to the different length scales, they are all in agreement when taking into account the uncertainties and each consistently recovers similar merger rates and perturbations at both the 10\msun and 35\msun peaks. In Figure \ref{fig:peak_sigs} we plot the posterior distribution of the perturbation function sliced at the approximate masses of the three, ($f(m_1\,=\,7.5\,\msun)$, $f(m_1\,=\,10\,\msun)$ and $f(m_1\,=\,35\,\msun)$). We find similar posteriors on the perturbation at these three mass regions, across the models varying the number of knots and the spline prior width. We calculate the percentile where $f=0$ falls in the posterior distribution for each of these three cuts, which would be near 50\% in the presence of no deviations to the power law or equivalently for draws from the spline model prior. The percentiles of zero perturbation for the combined model with the conservative (wide) priors are 70.8\% (72.6\%), 19.4\% (13.1\%), and 0.8\% (2.5\%) at $7.5\msun$, $10\msun$, and $35\msun$, respectively.

\begin{figure*}[t]
    \centering
    \includegraphics[width=\textwidth]{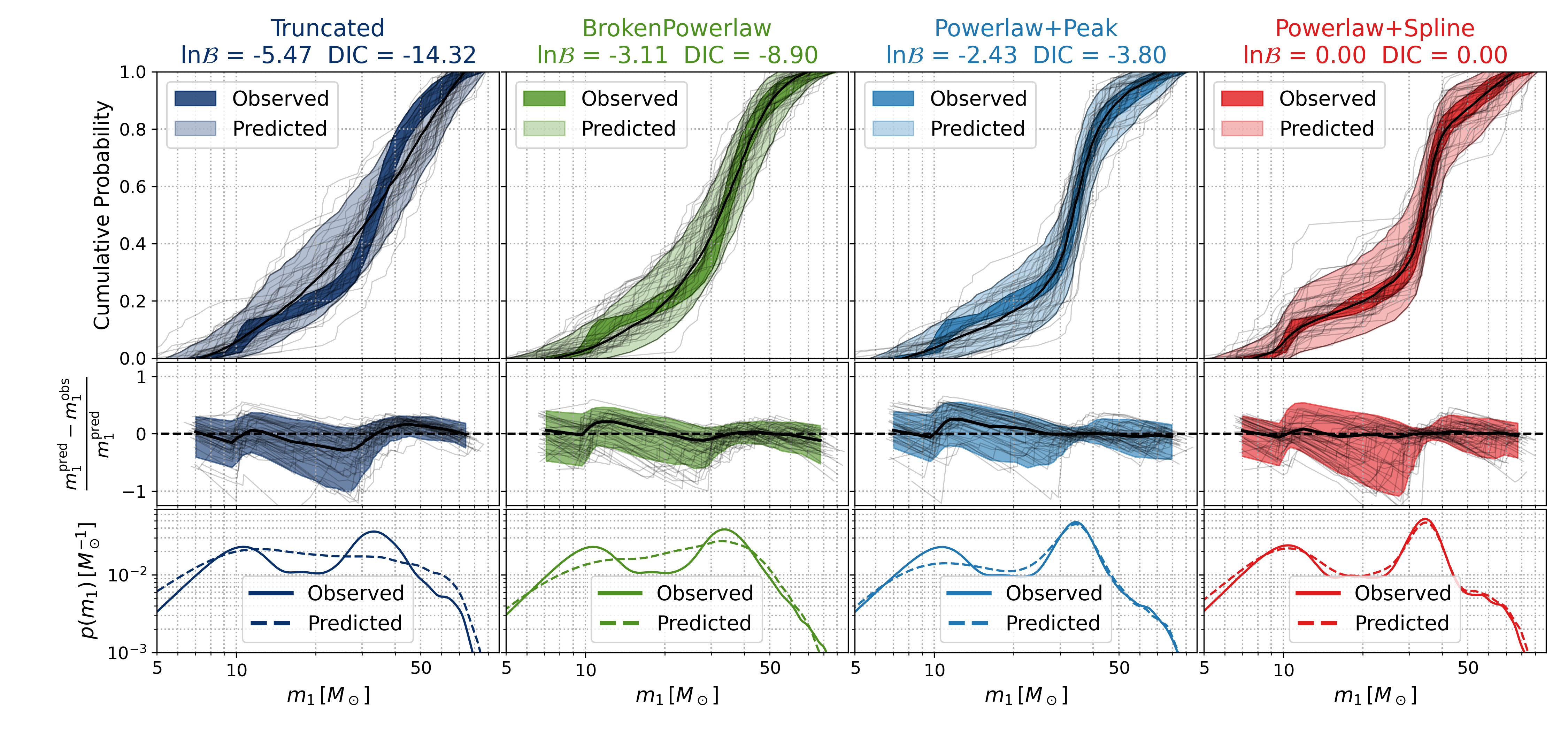}
    \caption{Posterior predictive checks for three of the parametric models used in \citet{o3a_pop} and the spline perturbation model of this work. We show the spline model result with the highest marginal likelihood which was the 15 knot and $\sigma_\mathrm{knot}=1$ model. The observed and predicted values for primary masses are generated by re-weighting either the injection set or the set of posterior samples for each BBH analyzed, for 500 draws from each models inferred posterior on the hyper-parameters, and then drawing 46 values from the re-weighted injections and a single fair draw from each of the 46 event re-weighted posterior samples. The top panel shows the CDF generated from these sets observed and predicted events for each of the 4 models, with the 90\% credible levels enclosed by the bands, the median in dark black, and the thin black lines showing 50 of the 500 sets of 46 predicted events. The middle row uses the same set of predicted and observed events and the y-axis shows the relative error in predicted to observed mass ($(m_1^\mathrm{pred}-m_1^\mathrm{obs})/m_1^\mathrm{pred}$) as a function of $m_1^{\mathrm{obs}}$. The last row of plots shows the PDF of the top row averaged over the 500 draws from the posterior on the hyper-parameters for both sets of events.}
    \label{fig:ppc}
\end{figure*}

The presence of the $\sim35\msun$ feature was previously found and reported in \citet{o3a_pop} as being either a peak (likely due to the PPISN pileup \citep{Talbot_2018}) or a break to a steeper power law. The \textsc{Powerlaw+Peak} model returned the highest marginal likelihood of parametric mass models considered in \citet{o3a_pop}, but was only favored with roughly 3:1 odds. Due to the inherent nature of the spline perturbation model, we would be more likely to find features that look like peaks rather than a power law break in the distribution. We additionally fit for spline perturbations on top of the \textsc{BrokenPowerlaw} model, which found little to no support for two different power law slopes and recovered a nearly identical primary mass distribution to what was found when modulating a single power law. The low mass feature recovered by our spline model was not identified in \citet{o3a_pop} because the models considered there did not have the flexibility to fit such features at low mass: they only include a smooth rise to the low-mass end of the power-law. This could be explained by additional structure that cannot be described by a smooth rise to a power law, coming from the upper edge of the proposed neutron star black hole mass gap. The flexibility of our semi-parametric approach enables us to find additional structure in the astrophysical mass distribution beyond what can be found with simpler toy models. While the spline perturbation model clearly finds structure beyond the power law around $\sim 35 \msun$ in the analyzed catalog of 46 BBH mergers, we cannot say for certain that the low mass feature is inherent to the astrophysical mass distribution. There is still the possibility that our model could be latching onto fluctuations in our data due to small number statistics. This possibility is reflected in the percentiles in Figure \ref{fig:peak_sigs}. The perturbation function at $7.5 \msun$ and $10 \msun$ does not rule out $f=0$ at high credibility regardless of prior choices in the spline model, while in contrast, the perturbation at $35 \msun$ rules out $f(35\msun)=0$ at 97-99.4\% credibility across each variation of spline model used. We investigate the possibility that these subsequent deviations from a power law could appear due to our model's systematics in Appendix \ref{sec:corrs}. We report no signs of correlations between the successive perturbations, which would be expected if the spline function was imposing biases onto our inferred perturbations. With future larger catalogs of gravitational-wave sources, we will be able to further resolve these low-mass features to determine if they are indeed present in the astrophysical mass distribution or a reflection of the current small catalog size.

\subsection{Posterior Predictive Checks}
With the large flexibility that comes from taking non-parametric approaches to modeling, one must be careful in validating inferences, especially in cases with small numbers of observations. One way we evaluate our semi-parametric model is through posterior predictive checks. We employ the same injection set used to estimate our search's selection effects mentioned in Appendix \ref{sec:hierarchical_inference} to create mock detected populations under a given population described by a posterior on hyper-parameters. We then compare these mock populations to the observed population. To do this, we first re-weight the injections to our inferred population for $N_\mathrm{draw}$ draws of hyper-parameters from our population posterior, then take $N_\mathrm{obs}$ (46 for this BBH-only analysis) draws for each of the re-weightings. This generates $N_\mathrm{draw}$ sets of $N_\mathrm{obs}$ ``Predicted'' observations for a given population inference. Next, we re-weight the individual event posterior samples to the same inferred population for $N_\mathrm{draw}$ draws of hyper-parameters from the posterior. For each draw of hyper-parameters, we take a fair draw from each re-weighted event posterior to generate our corresponding $N_\mathrm{draw}$ sets of $N_\mathrm{obs}$ ``Observed'' observations for a given population inference. From this procedure we generated 500 sets of 46 ``Observed" and ``Predicted" catalogs, which we compare to each other in Figure \ref{fig:ppc} to confirm that our inferred population model predictions are consistent with the observations. We show the cumulative probability as a function of observed primary mass in the top row, the relative error in predicted primary masses to observed in the second row, and the last row shows the observed and predicted primary mass distributions averaged over all of the hyper-parameters inferred in our posterior. The colored bands in the top row of Figure \ref{fig:ppc} show that a model is inconsistent with observations when the dark ``Observed" band extends outside of the lighter ``Predicted" band. We see this behaviour at $\sim40$--$50\msun$ for the \textsc{Truncated} model which illustrates a conclusion from \citet{o3a_pop}: the \textsc{Truncated} model is inconsistent with the mergers in GTWC-2. The spline perturbation model is the only primary mass model that recovers both the low and high mass features seen in the observed distribution, but it does exhibit more uncertainty than other models in the regions between the $\sim10\msun$ and $\sim35\msun$ peak. When considering possible fluctuations due to small number statistics, the observations at low mass are still consistent with both the \textsc{Powerlaw+Peak} and \textsc{BrokenPowerlaw} models.

\subsection{Astrophysical Implications}

The BBH mass distribution is particularly well-suited to answering a wide range of astrophysical questions. In particular, the masses of detected events are relatively well-measured, and different channels of BBH formation result in different mass distributions \citep[e.g.][]{Zevin_2017}, implying that the formation history is encoded within these distributions. With the tens of detected events available now, disentangling the overlapping sub-populations in the full population (if they exist) is a challenge. Our perturbation model can be used to see if or where a distribution describing a single (dominant) sub-population or formation channel may fail to fully fit the data, which would provide evidence that there may be non-negligible contributions due to additional formation channels. The hints of structure we see at the low end of the mass distribution  could point to such a superposition of multiple formation channels. Another factor that can affect the mass distribution is the physics of PISN or PPISN. Stellar evolution models describing mass loss and PPISN have uncertainties that can drastically change predictions on the masses at which the PPISN and PISN play a role. For example, choices in nuclear reaction rates within stellar cores can affect the BBH mass distribution \citep{Farmer_2019, Farmer_2020}. Our spline model enables us to measure these imprints of BBH formation in the observed distribution without enforcing the specific distribution shapes that are inherent to parametric models. Our findings corroborate the existence of a PISN ``pile-up" in the 30 - 40 $\msun$ range, and we infer its shape without assuming a simple functional form. With more GW detections, further resolution of this peak with the spline model could offer insights into supernova physics.


\section{Conclusions}\label{sec:conclusion}

Accurate estimation of the BBH mass distribution is paramount to getting accurate estimates of merger rates, the GW stochastic background, and false alarm rates for potential new triggers. Low-dimensional parametric models have the advantage of being easily interpreted but are limited in their flexibility and subject to a-priori expectations. We presented a semi-parametric approach to modeling the primary mass distribution of BBH mergers, using cubic splines that modulate a power law. We show that our flexible semi-parametric approach, when applied to the BBHs in GWTC-2, consistently recovers the previously reported excess of observed mergers near $\sim35\msun$, and shows potential signs of additional features in the low-mass end of the BBH distribution. These low-mass features beyond the power law structure, correspond with similar features found in the chirp mass distribution using a separate non-parametric approach based on a flexible Gaussian Mixture model \citep{Tiwari_2021_b}. We show through posterior predictive checks that the spline model is at least as good as the \textsc{Powerlaw+Peak} model at fitting the high mass structure in our catalog while having the flexibility beyond a smooth rise into a power law to capture the apparent excess at $10\msun$. Structure in the mass distribution that could arise from many different astrophysical phenomena but if we are able to confidently identify either a high-mass cutoff or pileup in the mass distribution it is likely to be related to effects of PISN or PPISN. These two features can both be used as calibrated mass-scales to measure a redshift-luminosity-distance relation with which it is possible to infer cosmological parameters with \citep{Farr_2019HUB}.

Our semi-parametric approach has advantages compared to other fully non-parametric approaches modeling the BBH mass distribution \citep{Mandel_2016, Tiwari_2021_a, Tiwari_2021_b}. The semi-parametric approach leverages the information learned from the parametric models to explain the majority of the structure, while reserving the flexibility to see where observations may start to diverge from previous inferences. This same method of applying cubic spline perturbations to simpler population models can be used on any of the other commonly modeled population distributions such as the mass ratio, spins or redshift evolution. With the relatively small catalog sizes currently available, structure, if present, would likely only appear in the best measured parameters. In future work, we plan to extend this method to incorporate multi-dimensional splines, that could uncover correlations between different parameters such as peaks in the mass distribution associated with a high spin magnitude. Such correlations would be a tell-tale sign of hierarchical mergers, for example \citep{Gerosa:2017kvu, Fishbach_2017, Kimball_genealogy, Doctor_2020, doctor2021black}. Future work could also extend this method to more than just the BBH mass distribution, and allow for adaptive resolution splines that allow the knot locations to vary \citep{Edelman_2021}. With additional observations of GWs associated with BNSs and NSBHs \citep{NSBH_discovery} our spline perturbation model is well suited to model the joint mass distribution of all GW-observed compact objects. This would complement the parametric model of \citep{Fishbach_2020_mm}, giving insights on the structure (or lack there of) of the ``lower mass gap" that may exist between the heaviest neutron stars and lightest black holes. 

\section{Acknowledgements}\label{sec:acknowledments}

We thank Will Farr for countless useful discussions related to this work and hierarchical modeling. This research has made use of data, software and/or web tools obtained from the Gravitational Wave Open Science Center (\url{https://www.gw-openscience.org/}), a service of LIGO Laboratory, the LIGO Scientific Collaboration and the Virgo Collaboration. This work benefited from access to the University of Oregon high performance computer, Talapas.  This material is based upon work supported in part by the National Science Foundation under Grant PHY-1807046 and work supported by NSF’s LIGO Laboratory which is a major facility fully funded by the National Science Foundation.

\software{
\textsc{Astropy}~\citep{2018AJ....156..123A},
\textsc{NumPy}~\citep{harris2020array},
\textsc{SciPy}~\citep{2020SciPy-NMeth},
\textsc{Matplotlib}~\citep{Hunter:2007},
\textsc{bilby}~\citep{Ashton_2019},
\textsc{GWPopulation}~\citep{Talbot_2019}
}

\bibliography{references}{}
\bibliographystyle{aasjournal}

\appendix
\section{Hierarchical Inference} \label{sec:hierarchical_inference}

\begin{table*}[ht!]
\centering
\begin{tabular}{|l|l|l|l|}
\hline
\textbf{Model} & \textbf{Parameter} & \textbf{Description} & \textbf{Prior} \\ \hline \hline
\multicolumn{4}{|c|}{\textbf{Primary Mass Model Parameters}} \\ \hline
\textsc{Truncated} & $\alpha$ & slope of the power law & U(-4, 12) \\ \cline{2-4} 
 & $m_\mathrm{min}$ & minimum mass cutoff & U(2\msun, 10\msun) \\\cline{2-4}
 & $m_\mathrm{max}$ & maximum mass cutoff & U(35\msun, 100\msun) \\ \cline{2-4}
 & $\delta_m$ & low-mass smoothing from [$m_\mathrm{min}$, $m_\mathrm{min} + \delta_m$] & U(0\msun, 10\msun) \\ \hline
\textsc{BrokenPowerLaw} & $\alpha_1$ & slope of first power law & U(-4, 12) \\ \cline{2-4}
 & $\alpha_2$ & slope of second power law & U(-4, 12) \\ \cline{2-4} 
 & b & fraction between $m_\mathrm{min}$ and $m_\mathrm{max}$ where the power law break lies & U(0, 1) \\ \cline{2-4} 
 & $m_\mathrm{min}$ & minimum mass cutoff & U(2\msun, 10\msun) \\ \cline{2-4}
 & $m_\mathrm{max}$ & maximum mass cutoff & U(50\msun, 100\msun) \\ \cline{2-4}
  & $\delta_m$ & low-mass smoothing from [$m_\mathrm{min}$, $m_\mathrm{min} + \delta_m$] & U(0\msun, 10\msun) \\ \hline
\textsc{PowerLaw+Peak} & $\alpha$ & slope of the power law & U(-4, 12) \\ \cline{2-4}  
 & $m_\mathrm{min}$ & minimum mass cutoff & U(2\msun, 10\msun) \\ \cline{2-4} 
 & $m_\mathrm{max}$ & maximum mass cutoff & U(30\msun, 100\msun) \\ \cline{2-4} 
 & $\mu_p$ & location in mass space where the Gaussian peak lies & U(20\msun, 70\msun) \\ \cline{2-4} 
 & $\sigma_p$ & width of the Gaussian peak & U(0.4\msun, 10\msun) \\ \cline{2-4} 
 & $\lambda_p$ & fraction of BBH in the Gaussian component & U(0, 1) \\ \cline{2-4}
  & $\delta_m$ & low-mass smoothing from [$m_\mathrm{min}$, $m_\mathrm{min} + \delta_m$] & U(0\msun, 10\msun) \\ \hline 
  \textsc{PowerLaw+MultiPeak} & $\alpha$ & slope of the power law & U(-4, 12) \\ \cline{2-4}  
 & $m_\mathrm{min}$ & minimum mass cutoff & U(2\msun, 10\msun) \\ \cline{2-4} 
 & $m_\mathrm{max}$ & maximum mass cutoff & U(30\msun, 100\msun) \\ \cline{2-4} 
 & $\mu_{p,1}$ & location in mass space where the first Gaussian peak lies & U(5\msun, 40\msun) \\ \cline{2-4} 
 & $\sigma_{p,1}$ & width of the first Gaussian peak & U(0.4\msun, 10\msun) \\ \cline{2-4} 
 & $\lambda_{p,1}$ & fraction of BBH in the first Gaussian component & U(0, 1) \\ \cline{2-4}
 & $\mu_{p,2}$ & location in mass space where the second Gaussian peak lies & U(20\msun, 70\msun) \\ \cline{2-4} 
 & $\sigma_{p,2}$ & width of the second Gaussian peak & U(0.4\msun, 10\msun) \\ \cline{2-4} 
 & $\lambda_{p,2}$ & fraction of BBH in the second Gaussian component & U(0, 1) \\ \cline{2-4}
 & $\delta_m$ & low-mass smoothing from [$m_\mathrm{min}$, $m_\mathrm{min} + \delta_m$] & U(0\msun, 10\msun) \\ \hline \hline
\multicolumn{4}{|c|}{\textbf{Mass Ratio Model Parameters}} \\ \hline
\textsc{PowerLawMassRatio} & $\beta_q$ & slope of the mass ratio power law & U(-4, 12) \\ \hline \hline
\multicolumn{4}{|c|}{\textbf{Redshift Evolution Model Parameters}} \\ \hline
\textsc{PowerLawRedshift} & $\lambda$ & slope of redshift evolution power law $(1+z)^\lambda$ & U(-2, 6) \\ \hline \hline
\multicolumn{4}{|c|}{\textbf{Spline Perturbation Model Parameters}} \\ \hline
\textsc{Cubic Spline} & $\{m_n\}$ & location in primary mass of the n spline interpolant knots & FIXED \\ \cline{2-4} 
 & $\{f_n\}$ & y-value of the spline interpolant knots & $N(\mu=0, \sigma=1)$ \\ \hline
\end{tabular} 
\caption{This table enumerates all the hyper-parameters, their descriptions, and chosen priors for this work for each respective population model we use. The \textsc{Truncated} model is extended from the version used in \citet{o3a_pop} to have the option of a low-mass taper of the same form as the \textsc{Powerlaw+Peak} model. Note that we do not describe a spin population model in this table since in this work we are not inferring a hyper-prior on the spins and instead assume they are described by the default (uniform in component magnitudes, isotropic in orientations) parameter estimation prior used to produce \citet{gwtc2}}. 
\label{tab:model_params}
\end{table*}
We use hierarchical Bayesian inference to simultaneously infer the population distributions of the primary masses ($m_1$), mass ratios ($q$) and the redshifts ($z$) of observed BBHs. For a set of hyper-parameters, $\Lambda$, and local ($z=0$) merger rate density (units of mergers per co-moving volume per time), $\mathcal{R}_0$, we write the overall number density of BBH mergers in the universe as: 

\begin{equation} \label{number_density}
     \frac{dN(m_1, q, z | \mathcal{R}_0, \Lambda)}{dm_1dqdz} = \frac{dV_c}{dz}\bigg(\frac{T_\mathrm{obs}}{1+z}\bigg) \frac{d\mathcal{R}(m_1, q, z | \mathcal{R}_0, \Lambda)}{dm_1dq} = \mathcal{R}_0 p(m_1 | \Lambda) p(q | m_1, \Lambda) p(z | \Lambda)
\end{equation}

\noindent
Above, $dV_c$ is the co-moving volume element \citep{hogg_cosmo} and $T_\mathrm{obs}$, the observing time period that produced the catalog with the related factor of $1+z$ converting this detector-frame time to source-frame. We assume a LambdaCDM cosmology using the cosmological parameters from \citet{Planck2015}. The redshift evolution of the merger rate follows $p(z|\lambda) \propto \frac{dV_c}{dz}\frac{1}{1+z}(1+z)^\lambda$. Integrating equation \ref{number_density} across all masses, and up to some redshift, $z_m$, returns the total number of BBH mergers in the universe out to that redshift. Let $\{d_i\}$ represent a set of data from $N_\mathrm{obs}$ observed gravitational waves associated with BBH mergers. We model the merger rate as an inhomogenous point process and when imposing a log-uniform prior on the merger rate, we can marginalize over the merger rate to get the posterior distribution of our hyper-parameters, $\Lambda$ \citep{Mandel_2019, Vitale_2021}.

\begin{equation} \label{posterior}
    p\left(\Lambda | \{d_i\} \right) \propto \frac{p(\Lambda)}{\xi(\Lambda)^{N_\mathrm{obs}}} \prod_{i=1}^{N_\mathrm{obs}} \Bigg[ \int \mathcal{L}\left(d_i | m_1^i, q^i, z^i \right) p(m_1 | \Lambda) p(q | m_1, \Lambda) p(z | \Lambda) dm_1 dq dz \Bigg],
\end{equation}

\noindent
where, $\mathcal{L}(d_i|m_1, q, z)$, is the single-event likelihood function from each events original analysis, and $\xi(\Lambda)$ is the detection efficiency given a population distribution described by $\Lambda$. The procedure for calculating $\xi(\Lambda)$ is described in more detail below. The LVC reports out posterior samples for each observed event, with which we can use importance sampling to estimate the integrals above in equation \ref{posterior}. We replace the integrals with ensemble averages over $K_i$ posterior samples associated with each event in the catalog:

\begin{equation}
 p\left(\Lambda | \{d_i\}\right) \propto p(\Lambda) \prod_{i=1}^{N_\mathrm{obs}} \bigg[ \frac{1}{K_i} \sum_{j=1}^{K_i} \frac{p(m_1^{i,j}|\Lambda)p(q^{i,j}|\Lambda)p(z^{i,j}|\Lambda)}{\pi(m_1^{i,j}, q^{i,j}, z^{i,j})} \bigg]
\end{equation}

\noindent 
Here $j$ indexes the $K_i$ posterior samples from each event and $\pi(m_1, q, z)$ is the default prior used by parameter estimations that produced the posterior samples for each event. In the analyses of GWTC-2 the default prior used is uniform in detector frame masses and Euclidean volume. The corresponding prior evaluated in source frame masses and redshift is $\pi(m_1, q, z) \propto m_1(1+z)^2 D_L^2(z) \frac{dD_L}{dz}$, where $D_L$ is the luminosity distance. 

To carefully incorporate selection effects to our model we need to quantify the detection efficiency, $\xi(\Lambda)$, of the search pipelines that were used to create GWTC-2, at a given population distribution described by $\Lambda$.
 
\begin{equation} \label{det_eff}
     \xi(\Lambda) = \int dm_1 dq dz P_\mathrm{det}(m_1, q, z)p(m_1 | \Lambda) p(q | m_1, \Lambda) p(z | \Lambda)
\end{equation}
 
\noindent The above integral is not tractable since there is no analytic prescription for $P_\mathrm{det}(m_1,q,z)$, the detection probability of an individual event. To estimate this integral we use a software injection campaign where GWs from a fixed, broad population of sources are simulated, put into real detector data, and then run through the same search pipelines that were used to produce the catalog we are analyzing \footnote{For O3a we used the injection sets used by \citet{o3a_pop}, which can be found at https://dcc.ligo.org/LIGO-P2000217/public. For O1/O2 we used the mock injection sets used by \citet{o1o2_pop} which can be found at https://dcc.ligo.org/LIGO-P2000434/public}. With these search results in hand, we use importance sampling to evaluate the integral in equation \ref{det_eff}:

\begin{equation} \label{xi}
     \xi(\Lambda) = \frac{1}{N_\mathrm{inj}} \sum_{i=1}^{N_\mathrm{found}} \frac{p(m_1^i | \Lambda) p(q^i | m_1, \Lambda) p(z^i | \Lambda)}{p_\mathrm{inj}(m_1^i, q^i, z^i)}
\end{equation}

\noindent
Where the sum indexes only over the $N_\mathrm{found}$ injections that were successfully detected out of $N_\mathrm{inj}$ total injections, and $p_\mathrm{inj}(m_1, q, z)$ is the reference distribution from which the injections were drawn. Additionally, we follow the procedure outlined in \citet{Farr_2019} to marginalize the uncertainty in our estimate of $\xi(\Lambda)$ due to a finite number of simulated events. We make the assumption that repeated sampling of $\xi(\Lambda)$ will follow a normal distribution with $\xi(\Lambda) \sim \mathcal{N}(\mu(\Lambda), \sigma(\Lambda))$, where the mean, $\mu$, is the estimate from equation \ref{xi}, while the variance, $\sigma^2$, is defined as:

\begin{equation}
    \sigma^2(\Lambda) \equiv \frac{\mu^2(\Lambda)}{N_\mathrm{eff}} \simeq \frac{1}{N^2_\mathrm{inj}} \sum_{i=1}^{N_\mathrm{found}} \bigg[\frac{p(m_1 | \Lambda) p(q | m_1, \Lambda) p(z | \Lambda)}{p_\mathrm{inj}(m_1, q, z)}\bigg]^2 - \frac{\mu^2(\Lambda)}{N_\mathrm{inj}}
\end{equation}

\noindent
Above we define $N_\mathrm{eff}$ as the effective number of independent draws contributing to the importance sampled estimate, in which we verify to be sufficiently high after re-weighting the injections to a given population (i.e. $N_\mathrm{eff} > 4N_\mathrm{obs}$). We write the hyper-posterior marginalized over the merger rate and uncertainty in estimation of $\xi(\Lambda)$, neglecting terms of $\mathcal{O}(N_\mathrm{eff}^{-2})$ or greater \citep{Farr_2019}, as:

\begin{equation}\label{importance-posterior}
    \log p\left(\Lambda | \{d_i\}\right) \propto \sum_{i=1}^{N_\mathrm{obs}} \log \bigg[ \frac{1}{K_i} \sum_{j=1}^{K_i} \frac{p(m_1^{i,j}|\Lambda)p(q^{i,j}|\Lambda)p(z^{i,j}|\Lambda)}{\pi(m_1^{i,j}, q^{i,j}, z^{i,j})} \bigg] -  \\
    N_\mathrm{obs} \log \mu + \frac{3N_\mathrm{obs} + N_\mathrm{obs}^2}{2N_\mathrm{eff}} + \mathcal{O}(N_\mathrm{eff}^{-2}).
\end{equation}

\noindent
We explicitly enumerate each of the models used in this work for $p(m_1|\Lambda)$, $p(q|m_1, \Lambda)$, and $p(z|\Lambda)$ along with their respective hyper-parameters and prior distributions in Table \ref{tab:model_params}. To calculate marginal likelihoods and draw samples of the hyper parameters from the hierarchical posterior distribution shown in equation \ref{importance-posterior}, we use the \textsc{Bilby} \citep{Ashton_2019, bilby_gwtc1} and \textsc{GWPopulation} \citep{Talbot_2019} Bayesian inference software libraries with the \textsc{Dynesty} dynamic nested sampling algorithm \citep{Speagle_2020}.

\begin{figure*}[ht!]
    \centering
    \includegraphics[width=\textwidth]{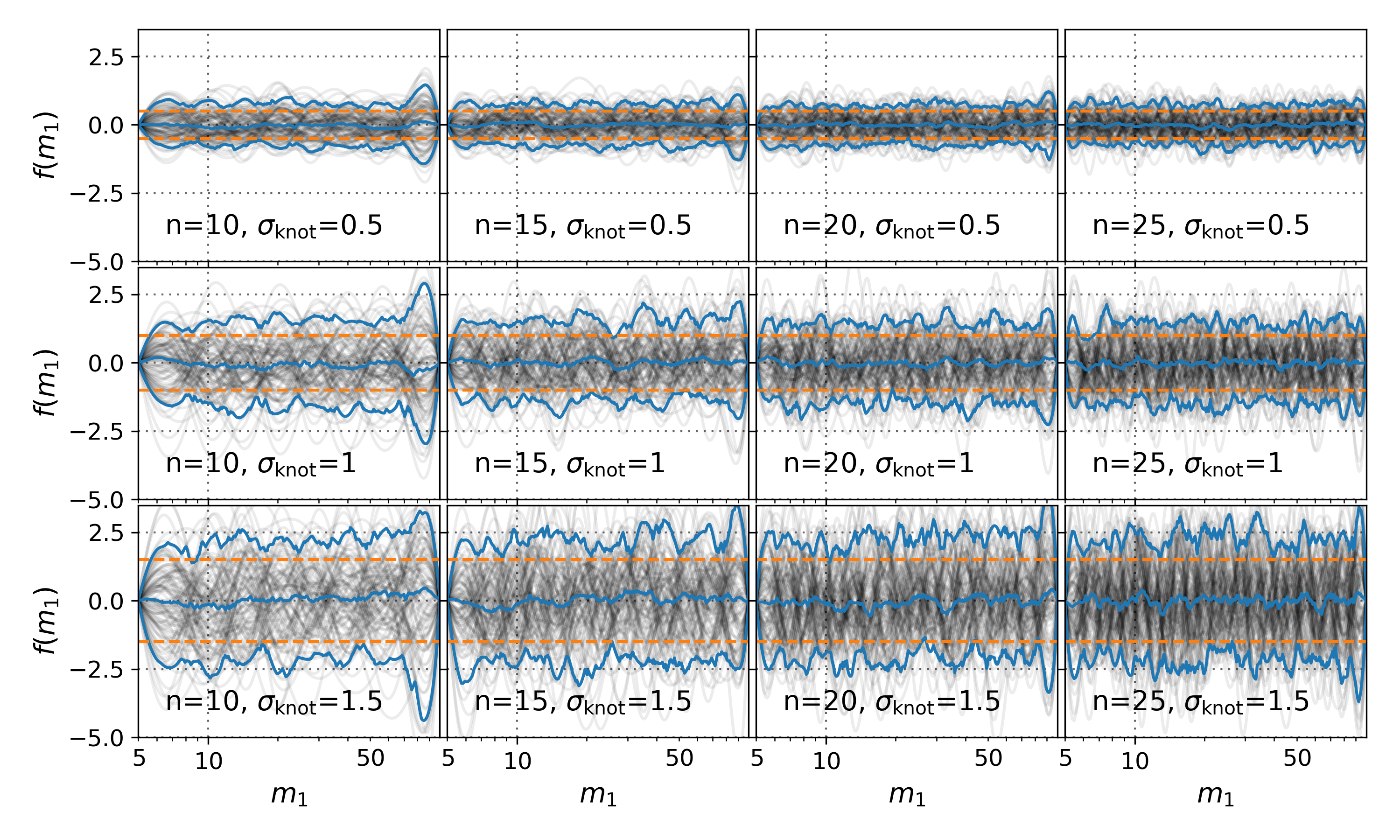}
    \caption{100 draws from the prior predictive distribution of the cubic spline function, $f(m_1)$, for different choices for the number of knots, n, and the width of the Gaussian priors on the knots, $\sigma_\mathrm{knot}$. The orange dashed line shows $\pm \sigma_\mathrm{knot}$, while the blue solid lines show the median and $1\sigma$ credible bounds of the draws from the prior.}
    \label{fig:prior_pred}
\end{figure*}

\section{Model Comparisons and Prior Specifications}\label{model-comparison}

\begin{figure*}[t]
    \centering
    \includegraphics[scale=0.75]{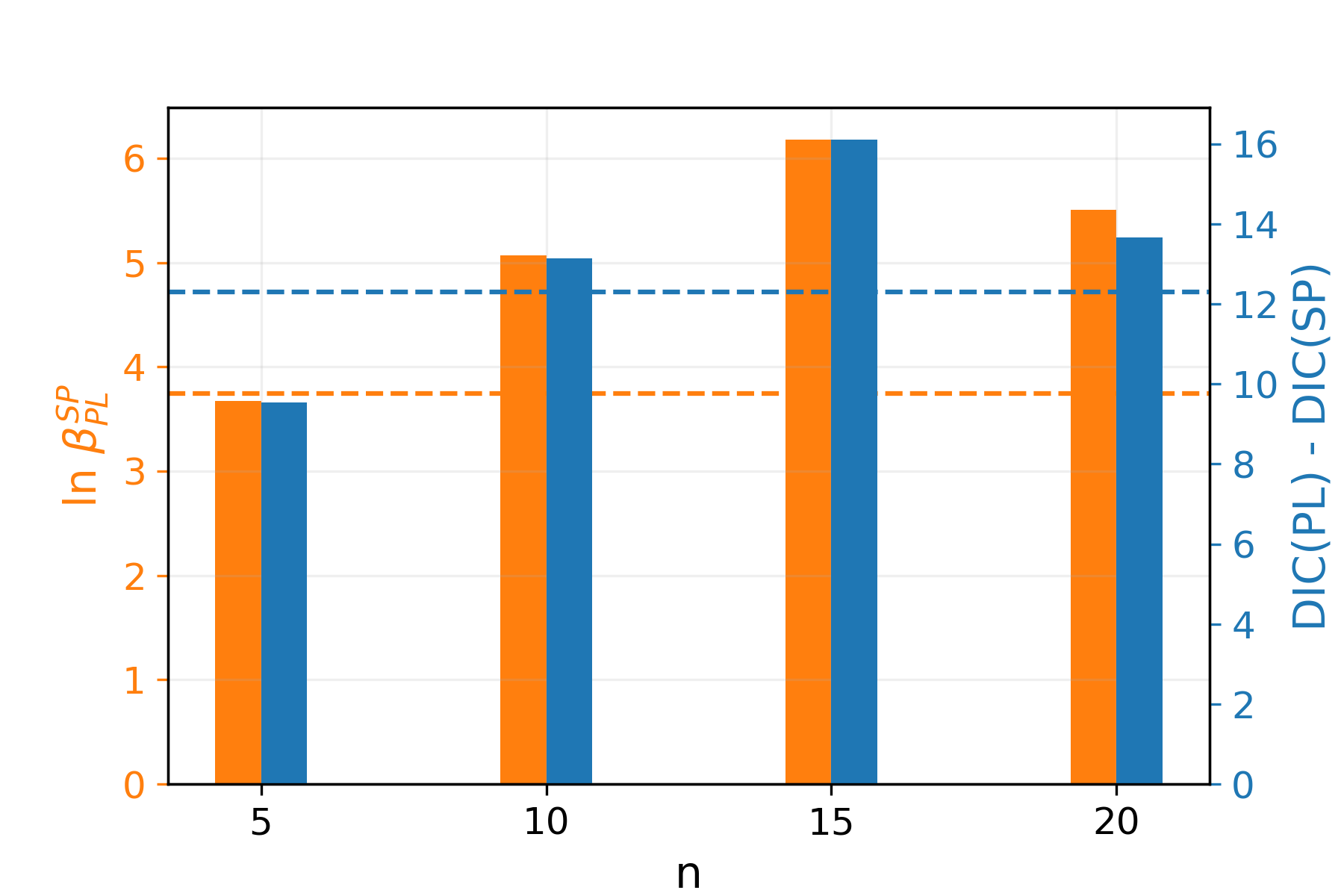}
    \caption{We show the two model comparison methods, Bayes factors (orange, left y-axis) and the DIC difference (blue, right y-axis), each comparing our spline perturbation model (denoted as SP) to the \textsc{Truncated} model (denoted as PL). The comparisons are calculated such that positive values of either metric denote the spline perturbation model being favored over the \textsc{Truncated}. Both values are shown for varying specifications for the spline prior. Along the x-axis we show different discrete choices (5, 10, 15, and 20) for the number of nodes, n. Each of these spline model analyses shown was performed with $\sigma_\mathrm{knot} = 1$. The horizontal dashed lines show the Bayes factor (orange) and DIC difference (blue) found when comparing \citet{o3a_pop}'s favorite mass model, \textsc{Powerlaw+Peak}, to the \textsc{Truncated} model.}
    \label{fig:node_comparison}
\end{figure*}

To compare competing population models in the aforementioned Bayesian framework we calculate two different measures of model goodness-of-fit, namely the marginal likelihoods ($\mathcal{Z}$) and deviance information criterion (DIC) \citep{DIC_2002}. The marginal likelihood for a given model is a constant that enforces that the posterior distribution is normalized (i.e. $\mathcal{Z} = \int d\Lambda p(\Lambda | \{d_i\})$), which has the property that it is higher for models that fit the data better or find higher likelihoods, while penalizing more complicated models by their prior volumes. As our semi-parametric approach has arbitrary prior choices one needs to make, this can significantly affect the marginal likelihoods inferred.We also calculate the DIC, a metric developed specifically for Bayesian hierarchical models \citep{DIC_2002}, which is less sensitive to the arbitrary prior choices for our semi-parametric model. While there are some limitations to the DIC \citep{DIC_12}, it provides a secondary metric to validate our model choices. The DIC is defined as:

\begin{equation}\label{dicdef}
    DIC = -2 \overline{\log(\mathcal{L})} + p_{D} \\
    = -2 \left( \overline{\log(\mathcal{L})} - \mathrm{var}(\log(\mathcal{L})) \right)
\end{equation}

\noindent With $\overline{\log\mathcal{L}}$ the mean log-likelihood, and $p_D$ the effective number of dimensions, defined as $p_D = \frac{1}{2}\mathrm{var}(-2\log\mathcal{L})$ with $\mathrm{var}(...)$ denoting the variance. Lower DICs indicate better models which, similarly to the marginal likelihood, favors models that find higher likelihoods while penalizing the more complicated models through the effective dimension term. We compare two models by calculating the ratio of their marginal likelihoods (i.e. Bayes Factors\footnote{The true ``Bayesian" way to compare models is using odds, which are Bayes factors multiplied by the ratio of prior odds of each model. Because we don't a priori have expectations of which population model would be more likely, we use Bayes factors which are odds ratios with equal prior weights for each model.}), defined as the ratio of each models marginal likelihoods. To compare DICs, we take the difference of two models values ($DIC\,\, \mathrm{dif} = DIC_A - DIC_B$) where positive differences indicate preference for model B, and negative differences indicate preference for model A. 

\begin{table*}
    \centering
    \begin{tabular}{|l|l|l|l|}
        \hline
        \bf{Model Name} & \bf{$\ln \mathcal{Z}$} & \bf{$\ln \mathcal{B}$} & \bf{$DIC\,\, \mathrm{dif}$} \\ \hline \hline
        Powerlaw+Spline ($n=15$, $\sigma_\mathrm{knot}=1$) & -347.10 & 0.00 & 0.00 \\ \hline
        Powerlaw+Spline ($n=20$, $\sigma_\mathrm{knot}=1$) & -347.77 & -0.67 & -2.43 \\ \hline
        Powerlaw+Spline ($n=15$, $\sigma_\mathrm{knot}=1$, $\delta_m = 0$) & -347.83 & -0.74 & -3.27 \\ \hline
        Powerlaw+Spline ($n=10$, $\sigma_\mathrm{knot}=1$) & -348.21 & -1.11 & -2.96 \\ \hline
        Powerlaw+Spline ($n=10$, $\sigma_\mathrm{knot}=1$, $\delta_m = 0$) & -348.24 & -1.14 & -3.25 \\ \hline
        Powerlaw+Spline ($n=20$, $\sigma_\mathrm{knot}=1$, $\delta_m = 0$) & -348.26 & -1.16 & -3.57 \\ \hline
        Powerlaw+Spline ($n=15$, $\sigma_\mathrm{knot}=2$) & -348.40 & -1.30 & -1.00 \\ \hline
        Powerlaw+Spline ($n=15$, $\sigma_\mathrm{knot}=2$, $\delta_m = 0$) & -348.53 & -1.43 & -2.85 \\ \hline
        Powerlaw+Spline ($n=20$, $\sigma_\mathrm{knot}=2$) & -348.62 & -1.52 & -3.00 \\ \hline
        Powerlaw+Spline ($n=20$, $\sigma_\mathrm{knot}=2$, $\delta_m = 0$) & -349.01 & -1.91 & -3.39 \\ \hline
        Powerlaw+MultiPeak & -349.27 & -2.18 & -6.87 \\ \hline
        Powerlaw+Spline ($n=10$, $\sigma_\mathrm{knot}=2$, $\delta_m = 0$) & -349.43 & -2.34 & -3.12 \\ \hline
        Powerlaw+MultiPeak ($\delta_m = 0$) & -349.44 & -2.35 & -6.27 \\ \hline
        Powerlaw+Peak & -349.53 & -2.43 & -3.80 \\ \hline
        Powerlaw+Spline ($n=10$, $\sigma_\mathrm{knot}=2$) & -349.66 & -2.57 & -3.22 \\ \hline
        Powerlaw+Peak ($\delta_m = 0$) & -349.70 & -2.60 & -4.96 \\ \hline
        Broken Powerlaw ($\delta_m = 0$) & -349.92 & -2.83 & -9.32 \\ \hline
        Broken Powerlaw & -350.21 & -3.11 & -8.90 \\ \hline
        Truncated & -352.57 & -5.47 & -14.32 \\ \hline
        Truncated ($\delta_m = 0$) & -353.27 & -6.18 &  -16.10 \\ \hline
    \end{tabular}
    \caption{Model comparison results, listing each model tested (semi-parametric spline model or parametric mass model from \citet{o3a_pop}) and their respective marginal likelihoods ($\mathcal{Z}$) along with $\ln \mathcal{B}$ and $DIC\,\, \mathrm{dif}$. Both comparison metrics for each of the listed models are relative to the ``best performing" model or the one with the highest (lowest) marginal likelihood (DIC), which, in both cases, was the Powerlaw+Spline ($n=15$, $\sigma_n=1$) model. We note that the \textsc{Powerlaw+MultiPeak} finds higher marginal likelihoods than the \textsc{Powerlaw+Peak} model which was not the case in \citet{o3a_pop}. This is because we used different priors for the \textsc{Powerlaw+MultiPeak} model that allowed for a peak at lower masses than the $\sim 35\,\msun$ peak instead of higher.}
    \label{tab:model_comparisons}
\end{table*}

\begin{figure*}[t]
    \centering
    \includegraphics[width=\textwidth]{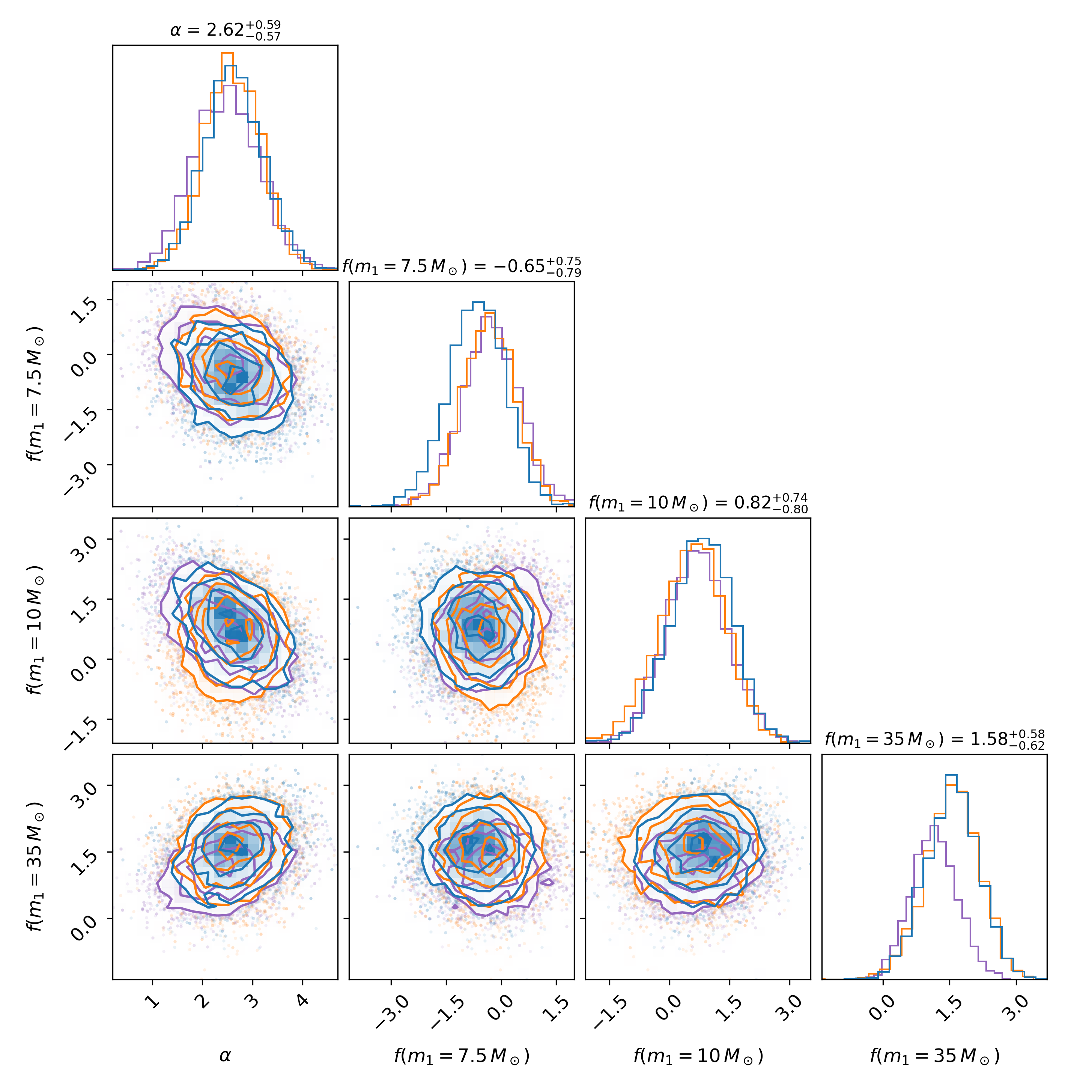}
    \caption{Corner plot that shows the posterior distribution on the power law slope, $\alpha$, and the height of the perturbation function, $f(m_1)$, sliced at the three masses of most significant deviation: $7.5\,\msun$, $10\,\msun$, and $35\,\msun$. We show the results for spline models with $\sigma_\mathrm{knot} = 1$ and 10 (purple), 15 (blue) and 20 (orange) nodes. The median and 90\% credible intervals quoted are for the 15 knot model.}
    \label{fig:peak_corr1}
\end{figure*}

\begin{figure*}[t]
    \centering
    \includegraphics[width=\textwidth]{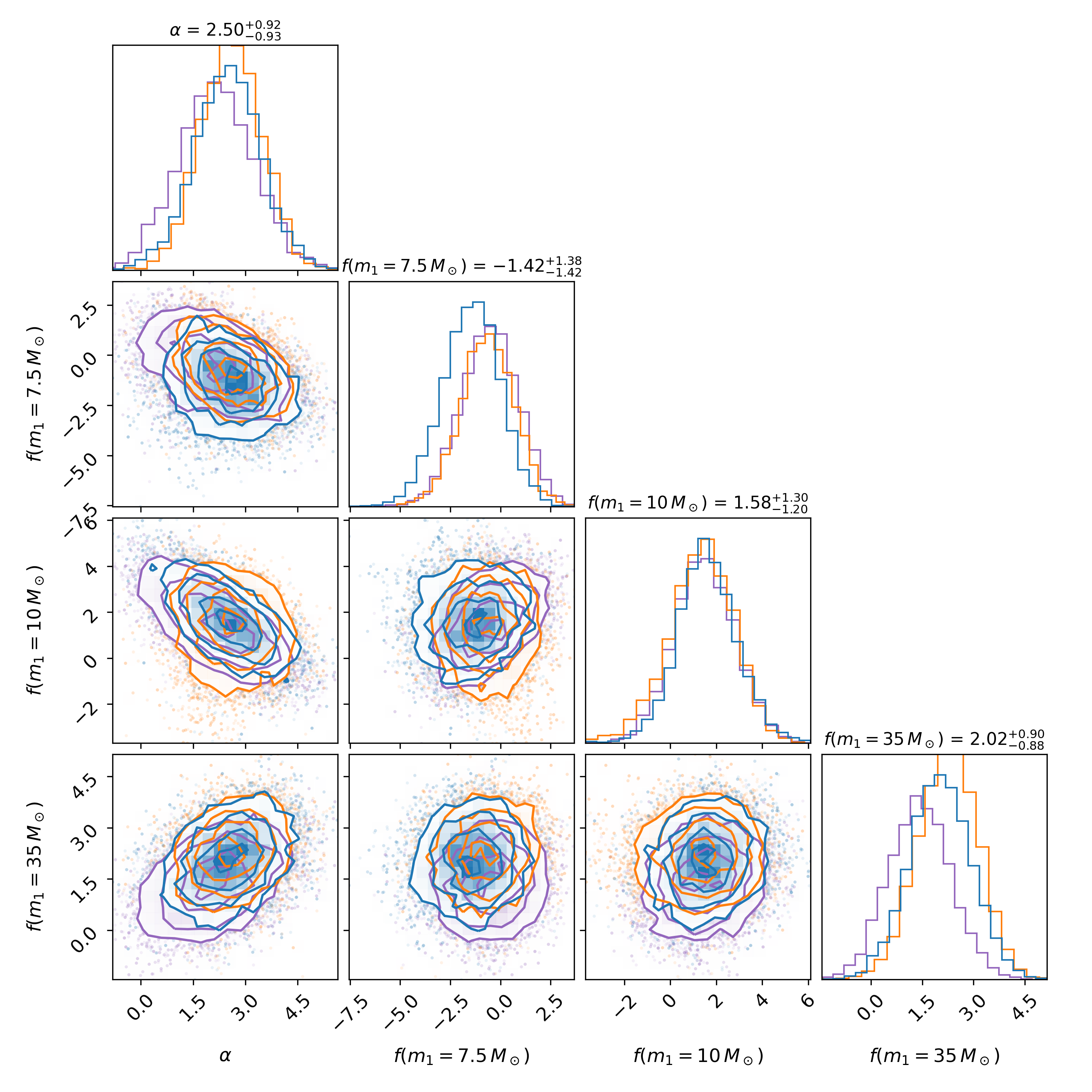}
    \caption{Corner plot that shows the posterior distribution on the power law slope, $\alpha$, and the height of the perturbation function, $f(m_1)$, sliced at the three masses of most significant deviation: $7.5\,\msun$, $10\,\msun$, and $35\,\msun$. We show the results for spline models with $\sigma_\mathrm{knot} = 2$ and 10 (purple), 15 (blue) and 20 (orange) nodes. The median and 90\% credible intervals quoted are for the 15 knot model.}
    \label{fig:peak_corr2}
\end{figure*}

We use these model comparisons to motivate a sensible choice for our spline models prior flexibility, namely the number of knots (n). Figure \ref{fig:prior_pred} shows changing prior widths on our knots only only effects magnitude of perturbations the spline is sensitive to. Additionally we see that as we add more knots, the model is free to fit sharper fluctuations. This flexibility comes with a penalty in our comparison metrics due to increased model complexity. Therefore, we would expect to see the the model comparisons increasingly favor our spline perturbation model as we increase our models complexity/flexibility up to a point where the penalty will start to overpower the higher likelihoods found with more flexibility. Figure \ref{fig:node_comparison} shows how the DIC differences and log Bayes factors ($\ln \mathcal{B}$) change when comparing the spline perturbation model to the \textsc{Truncated} model with different choices for $n$. We see that the comparisons favoring our spline model peaks around 15 knots, indicating that 15 knots is a good trade-off between our models flexibility and goodness of fit. We also report the marginal likelihoods and model comparisons (relative to the most preferred model) for each of the parametric primary mass models from \citet{o3a_pop} and various specifications of the spline model in table \ref{tab:model_comparisons}. From this table we see that the spline model is consistently favored despite our arbitrary model specifications, giving credence to the hypothesis that there are features in the data our semi-parametric method is capable of finding that previously used parametric mass models are not sensitive to. We do not use the comparisons in Table \ref{tab:model_comparisons} to determine the validity of the \textsc{Powerlaw+Spline} models over others, and further studies on simulated populations and the effect of small number statistics are needed to fully assess the significance and robustness of these features. However, as catalogs of BBH mergers increase in size, the impact of small number statistics will diminish.

\section{Correlations of Peaks} \label{sec:corrs}


We look for the effect of our spline function biasing the inferred perturbation function by plotting a corner plot of the value of $f(m_1)$ sliced through the masses that show the largest deviations. This is shown in Figure \ref{fig:peak_corr1} for the conservative knot prior and Figure \ref{fig:peak_corr2} for the wide knot prior. If the dip followed by the peak feature at 7.5 and 10 solar masses was found due to the nature of cubic splines we would expect to see correlations between the heights at these values which do not appear in either Figure \ref{fig:peak_corr1} or \ref{fig:peak_corr2}. Since we are fitting for the underlying power law model simultaneously with the perturbations we also might expect to see some correlations with the power law slope and the peaks. There are slight signs of an expected anti-correlation of the 10\msun peak height and the power law slope, and corresponding correlation of the 35\msun peak height with the slope. This happens due to the degeneracy between the parametric and non-parametric portions of our model. If the 10\msun peak is small, the power law slope becomes steeper so that the ``turnover" power law at low mass can contribute to fitting this over-density. With a steeper power law slope, the power law portion of the model under fits the excess at 35\msun, leading to a larger peak found in the perturbation. We also note these correlations with the power law slope are more apparent in the lower resolution spline models and under the wider prior on the knots. 

\end{document}